\documentclass{emulateapj}
\input{epsf}
\submitted{Accepted by ApJ on October 27, 2010}

\shorttitle{The inner structure and kinematics of the Sagittarius dwarf galaxy}

\shortauthors{E. L. {\L}okas et al.}

\begin{document}

\title{The inner structure and kinematics of the Sagittarius dwarf galaxy\\ as a product of tidal stirring}

\author{Ewa L. {\L}okas\altaffilmark{1}, Stelios Kazantzidis\altaffilmark{2}, Steven R. Majewski\altaffilmark{3},\\
David R. Law\altaffilmark{4}, Lucio Mayer\altaffilmark{5} and Peter M. Frinchaboy\altaffilmark{6}}

\altaffiltext{1}{Nicolaus Copernicus Astronomical Center, 00-716 Warsaw, Poland; lokas@camk.edu.pl}
\altaffiltext{2}{Center for Cosmology and Astro-Particle Physics; and Department of Physics; and Department of Astronomy,
    The Ohio State University, Columbus, OH 43210, USA; stelios@mps.ohio-state.edu}
\altaffiltext{3}{Department of Astronomy, University of Virginia, Charlottesville, VA 22904-4325, USA; srm4n@virginia.edu}
\altaffiltext{4}{Department of Physics and Astronomy, University of California, Los Angeles, CA 90095, USA; Hubble Fellow;
drlaw@astro.ucla.edu}
\altaffiltext{5}{Institute for Theoretical Physics, University of Z\"urich, CH-8057 Z\"urich, Switzerland;
        lucio@phys.ethz.ch}
\altaffiltext{6}{Department of Physics and Astronomy, Texas Christian University, Fort Worth, TX 76129, USA;
p.frinchaboy@tcu.edu}

\begin{abstract}

  The tidal stirring model envisions the formation of dwarf spheroidal
  (dSph) galaxies
  in the Local Group and similar environments
  via the tidal interaction of
  disky dwarf systems with a larger host galaxy like the Milky Way.
  These progenitor disks are embedded in extended dark
  halos and during the evolution both components suffer strong mass
  loss. In addition, the disks undergo the morphological
  transformation into spheroids and the transition from ordered to
  random motion of their stars. Using collisionless $N$-body simulations
  we construct a model for the nearby and highly elongated Sagittarius
  (Sgr) dSph galaxy within the framework of the tidal stirring
  scenario.  Constrained by the present orbit of the dwarf, which is
  fairly well known, the model suggests that in order to produce the
  majority of tidal debris observed as the Sgr stream, but not yet
  transform the core of the dwarf into a spherical shape, Sgr must
  have just passed the second pericenter of its current orbit around
  the Milky Way.  In the model, the stellar component of Sgr is still
  very elongated after the second pericenter and morphologically
  intermediate between the strong bar formed at the first pericenter
  and the almost spherical shape existing after the third pericenter.
  This is thus the first model of the evolution of the Sgr dwarf that
  accounts for its observed very elliptical shape. At the present time
  there is very little intrinsic rotation left and the velocity
  gradient detected along the major axis is almost entirely of tidal
  origin. We model the recently measured velocity dispersion profile
  for Sgr assuming that mass traces light and estimate its current
  total mass within 5 kpc to be $5.2 \times 10^8 M_{\odot}$.  To have
  this mass at present, the model requires that the initial virial
  mass of Sgr must have been as high as $1.6 \times 10^{10}
  M_{\odot}$, comparable to that of the Large Magellanic Cloud, which
  may serve as a suitable analog for the pre-interaction, Sgr
  progenitor.
\end{abstract}


\section{Introduction}

The tidal stirring scenario (Mayer et al. 2001; Klimentowski et al. 2007, 2009; Kazantzidis et al. 2010) proposes that
the population of dwarf spheroidal (dSph) satellite galaxies around the Milky Way and Andromeda galaxies were
transformed into their present morphology via tidal interaction of late type progenitors
--- initially small
disks embedded in extended dark matter halos ---
with their respective host.
Due to tidal forces the dwarfs are substantially
stripped of their mass and their stellar component undergoes strong
dynamical and morphological evolution. This evolution manifests itself in the morphological transformation of the disk into
a bar and then an
ellipsoid or spheroid, with the commensurate transition from ordered (rotation in the disk) to random motion of the stars.
In the
transitory bar-like stage the orbits are mostly radial and become more isotropic as the stellar shape evolves toward
spherical. The effectiveness of the tidal transformation depends crucially on the orbital parameters (the orbital time
and the pericenter distance) and also to some extent on the initial structure of the dwarf (Kazantzidis et al.
2010).

An interesting example of a dSph galaxy strongly affected by tidal forces from the Milky Way is the Sagittarius
(Sgr) dwarf discovered by Ibata et al. (1994). A convincing proof of such interaction was provided by the
identification within the 2MASS survey
of the Sgr stream, which spans a 360$^\circ$ great circle on the sky (Majewski et al. 2003).
Until now, modeling of the Sgr system by $N$-body simulations has focused
on reproducing the properties of the tidal stream,
especially the velocities, velocity dispersions, positions
and distances of the stellar debris in the leading and trailing arm (Johnston et
al. 1995, 1999; Helmi \& White 2001; Helmi 2004; Martinez-Delgado et al. 2004; Johnston et al. 2005; Law et al. 2005;
Fellhauer et al. 2006; Pe\~narrubia et al. 2010).
Only recently, however, have models been developed that are able simultaneously to satisfy all angular position,
distance, and radial velocity constraints on both the leading and trailing tidal streams
(Law et al. 2009; Law \& Majewski 2010).
Nevertheless, in most such numerical models the dwarf was initially assumed to be spherical and one-component, with
the distribution of stars and dark matter approximated by a single Plummer sphere.

However, it has been known since the time of the discovery of Sgr that the shape of the dwarf is strongly non-spherical.
Ibata et al. (1997) estimated the shape to be bar-like with axis ratios 3:1:1 and this was later confirmed by
observations of the distribution of Sgr M giant stars with the 2MASS survey (Majewski et al. 2003).
The elongation of the dwarf is usually interpreted as due to tidal deformation (e.g., Johnston et al. 1995).
However, if a satellite that is initially spherical
is required to survive long enough to produce the observed amount of tidal debris,
the observed degree of elongation would be difficult to obtain
through tides. Even at
pericenter, where tidal forces are the strongest, the surface density contours will likely be elongated only in the
outer part of the dwarf (where the transition to the tidal tails occurs)
but remain rather circular in the central
parts.

\begin{figure}
\begin{center}
    \leavevmode
    \epsfxsize=8.5cm
    \epsfbox[0 0 240 240]{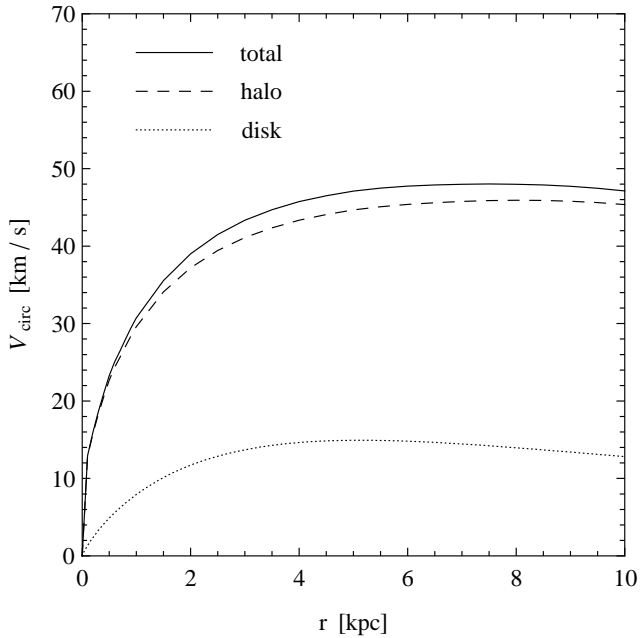}
\end{center}
\caption{The circular velocity curve of the simulated dwarf galaxy in the initial state. The dotted,
dashed and solid lines correspond, respectively, to the disk, the dark halo and the sum of the two.}
\label{circvel}
\end{figure}

\begin{figure}
\begin{center}
    \leavevmode
    \epsfxsize=8.5cm
    \epsfbox[0 30 400 390]{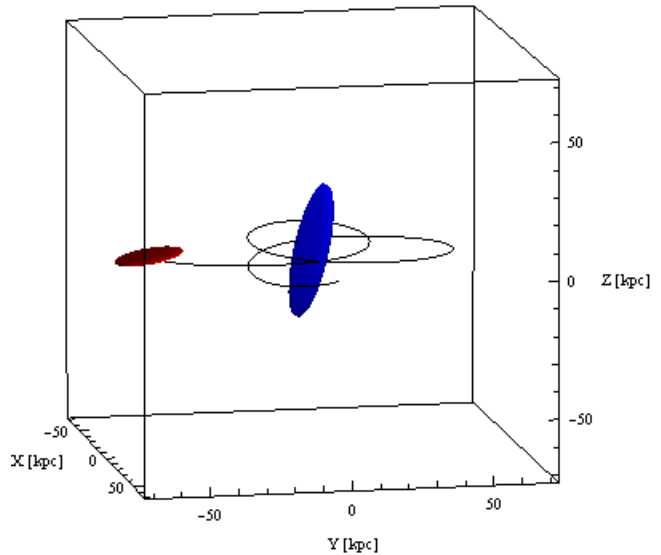}
\end{center}
\caption{A schematic view of the initial simulation setup. The dwarf galaxy, indicated by the smaller,
red disk on the left,
is located at the orbit apocenter and is inclined by 10$^\circ$ with respect to the orbital plane.
The bigger, blue disk
on the right shows the Milky Way disk. The orbit, plotted with the
solid black line, is coplanar with the $XY$ plane of the coordinate system and is inclined by 76$^\circ$
to the Galactic plane. The dwarf galaxy starts orbiting towards the positive $X$ coordinate and the dwarf
disk is prograde with respect to the orbit so that both the orbital motion of the dwarf and the rotation of the
stars in the dwarf are counterclockwise when viewed from the top of the figure. The Milky Way disk rotates
clockwise when viewed from the right side of the figure.}
\label{initial}
\end{figure}

In this work we demonstrate that the observed shape of the Sgr core can be reproduced if the stellar component was
initially in the form of a disk, slightly inclined with respect to the orbital plane. As such, the Sgr dwarf might have
previously resembled the Large Magellanic Cloud, a
conjecture that is actually consistent with other observations,
such as Sgr's chemistry (Chou et al. 2010) and star formation history (Siegel et al. 2007).
In our model, we embed the disk
in a massive dark matter halo, place it at an apocenter of a probable orbit of Sgr, determined from previous work,
and evolve the system for a few orbital times. The dwarf undergoes a characteristic
evolution on such a tight orbit: the stellar component forms a bar at the first pericenter that survives until
after the second. By that time a sufficient number of stars are stripped to reproduce
most of the observed Sgr stream (although see Correnti et al. 2010).
Unlike previous studies that focused on the Sgr tidal tails,
here we use constraints from the shape of the Sgr core to estimate the time Sgr has spent on its current orbit around
the Milky Way and the number of pericenters it has passed while on that orbit.

The paper is organized as follows. In Section 2 we describe the $N$-body simulations used in this study.
In the next section we
study the evolution of the simulated Sgr dwarf. Section 4 is devoted to the modeling of the stellar kinematics in the
simulated
Sgr system and a comparison to those in the real Sgr.
The discussion of the implications of our model follows in Section 5.

\section{The simulations}

We performed collisionless $N$-body simulations of rotationally supported, disky
dwarf galaxies orbiting inside a Milky Way-sized host.
All simulations were
carried out with the multistepping, parallel, tree $N$-body code PKDGRAV
(Stadel 2001).

Given existing uncertainties about the exact shape of the Milky Way
halo (i.e., whether it is prolate, oblate or triaxial; see, e.g., Helmi 2004;
Law et al. 2005, 2009; Law \& Majewski 2010) and the fact that our purpose
here is not to reproduce perfectly the shape and kinematics of the Sgr stream
but rather the shape and kinematics of the Sgr core, we decided to employ a simple Milky Way model with a
spherical dark matter halo.
Because the dominant aspect affecting the evolution of the core is the strength of the tidal force,
dictated primarily by the perigalacticon distance of the dwarf and the radial mass
distribution of the Milky Way but not its detailed shape, we do not expect
this simplification to affect our general results.

The adopted model is based on the
dynamical mass model A1 for the Milky Way from Klypin et al. (2002), which
consists of a Navarro et al. (1997, hereafter NFW) halo with a virial
mass of $M_{\rm vir} = 10^{12} M_{\odot}$ and concentration $c=12$,
a stellar disk with a mass of $M_{\rm D} = 4 \times 10^{10}
M_{\odot}$, scale length of $R_{\rm d} = 3.5$ kpc, and scale
height of $z_{\rm d} = 0.35$ kpc, as well as a bulge with a mass of $M_{\rm
b}=0.008 M_{\rm vir}$ and scale-length of $a_{\rm b} = 0.2 R_{\rm
d}$. The host galaxy model is live, which allows us to take into
account the effect of dynamical friction on the satellite.  The $N$-body realization of
the Milky Way model contains $N_{\rm h} = 10^{6}$ particles in the halo and
$N_{\rm d}=2 \times 10^{5}$ in the disk and bulge and is built using the
technique developed by Hernquist (1993). We used a fairly large gravitational softening for
the dark matter particles, $\epsilon=2$ kpc, to minimize spurious two-body heating between
massive halo particles and those of the disk in the dwarf. The softening in the stars of the
Milky Way was set to $\epsilon=100$ pc.

We employ the method of Widrow \& Dubinski (2005) to construct
numerical realizations of self-consistent, multi-component dwarf
galaxies built of exponential stellar disks embedded in NFW dark
matter halos. The Widrow \& Dubinski (2005) models represent
axisymmetric, equilibrium solutions to the coupled collisionless
Boltzmann and Poisson equations, and are thus ideal to confirm both
the tidal heating and the tidally-induced non-axisymmetric
instabilities that are required for the transformation of the dwarf.

\begin{figure}
\begin{center}
    \leavevmode
    \epsfxsize=8.5cm
    \epsfbox[0 20 290 590]{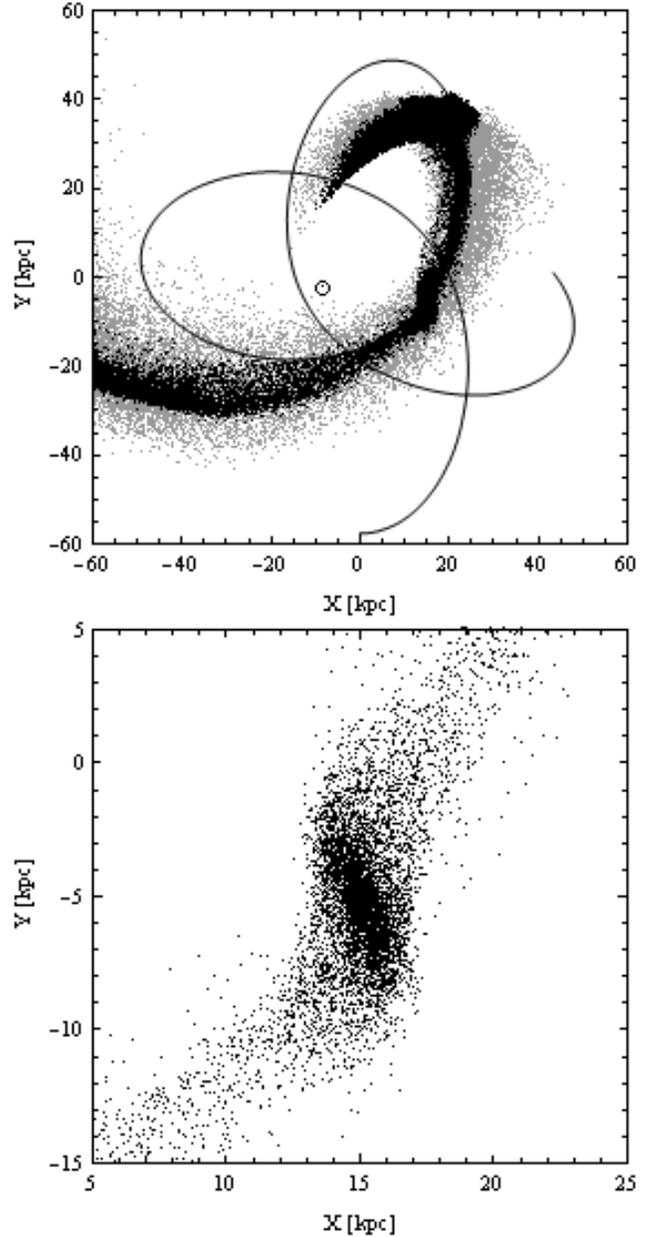}
\end{center}
\caption{({\it Upper panel\/})
Overview of the simulated Sgr projected onto the orbital plane. The gray dots show the dark matter particles,
the black ones the stars. The solid line plots the orbit of the galaxy during 2.5 Gyr of evolution with the starting
position at the bottom of the plot. The Milky Way disk intersects the orbital plane at $y=0$. The Sun symbol
shows the position of the Sun. ({\it Lower panel\/}) The magnified view of the stellar component of the simulated
Sgr.}
\label{overview}
\end{figure}

\begin{figure}
\begin{center}
    \leavevmode
    \epsfxsize=8.5cm
    \epsfbox[45 0 250 230]{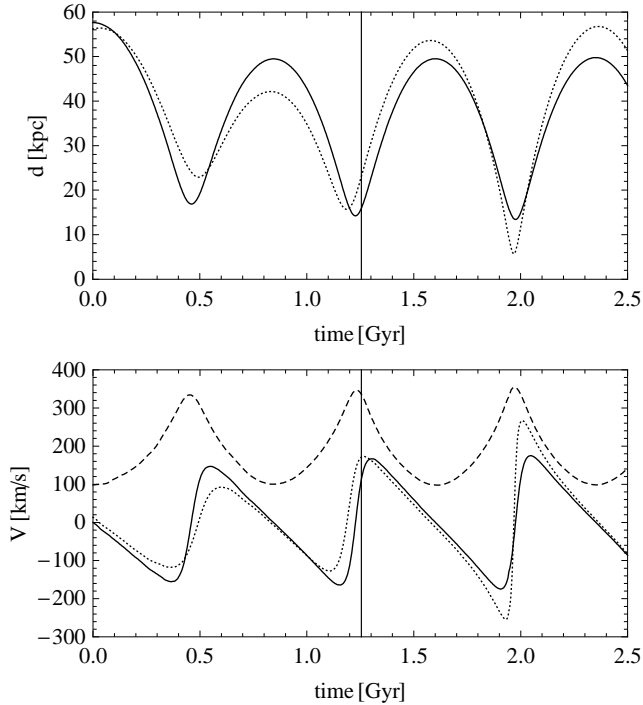}
\end{center}
\caption{({\it Top panel\/}) The distance of the Sgr center from the Galactic Center (GC; solid line)
and from the Sun (dotted line).
({\it Bottom panel\/}) The total velocity of the Sgr center with respect to the GC (dashed line) and the radial
velocities with
respect to the GC (solid line) and the Sun (dotted line). In both panels the vertical line indicates the present
position of Sgr after 1.255 Gyr of evolution, which corresponds to the output shown in Figure~\ref{overview}.}
\label{distvel}
\end{figure}

In the simulation described below that reproduced best the observed properties of Sgr,
the dwarf progenitor initially had a dark halo with a virial mass $M =
1.6 \times 10^{10} M_{\odot}$ and a concentration of $c=15$. The mass
and radial scale length of the disk were $3.2 \times 10^8 M_{\odot}$
and $R_{\rm d} = 2.3$ kpc, respectively. According to Mo et
al. (1998), this disk scale-length corresponds to an angular momentum
parameter $\lambda = J/G \sqrt{|E|/ M_{\rm vir}^5} = 0.08 $ (where $J$
and $E$ are the total halo angular momentum and energy, respectively),
which is typical of dwarf galaxies (Jimenez et al. 2003). We
note that this value of $R_{\rm d}$ is derived without considering the
effect of halo adiabatic contraction in response to the accretion
of baryons.
For the disk vertical structure we assumed an isothermal sheet with
scale height of $z_{\rm d}=0.3\,R_{\rm d}$, which is higher compared
to that appropriate for massive galaxies. Such choice is motivated by the
greater importance of turbulent motions in dwarf galaxies which
results in thicker systems (e.g., Schombert 2006). This disk thickness parameter
corresponds to the axis ratio $c/a = 0.18$ calculated from the moments of the
inertia tensor for all stars in the disk. The initial central value of the
velocity dispersion associated with the vertical structure of the disk was
$\sigma_z=19.5$ km s$^{-1}$. The decomposition
of the circular velocity curve into the different components of the
dwarf is shown in Figure~\ref{circvel}.

We sampled the dwarf galaxy with $10^{6}$
dark matter particles and $1.2 \times 10^{6}$ stellar disk
particles. The gravitational softening length was set to $\epsilon=60$ and
$15$~pc, for the dark matter and stellar particles, respectively.
Evolution of the dwarf model in isolation confirmed its stability
against bar formation for $10$~Gyr, and demonstrated the excellent
quality of the initial conditions
as well as adequate resolution of the
simulations.

The dwarf galaxy was evolved on an eccentric orbit with an initial
apocenter and pericenter of $r_{\rm a}=58$ kpc and $r_{\rm p}=17$ kpc,
respectively. We placed the dwarf initially at the apocenter and
followed its evolution for 2.5 Gyr, which corresponds to more than three
orbital periods. The orbit was inclined by 76$^\circ$ to the Galactic
plane and the dwarf galaxy disk was inclined by 10$^\circ$ with respect to
the orbital plane. The rotation in the disk was
prograde with respect to the orbital motion.
Our choice of the orbital parameters was motivated by the study of Law et al. (2005) for the spherical Milky Way
halo case (with small modifications due to the fact that we use a slightly different model of the Milky Way).
The inclination of the dwarf galaxy disk and the direction of its rotation were adjusted to reproduce the inclination
of the observed Sgr image with respect to the orbit and the observed rotation curve, respectively (see the next
sections). The initial simulation setup is illustrated in Figure~\ref{initial}.

We note that for such a massive dwarf
the initial dark halo is very extended, with a virial radius of $\sim
65$ kpc. Therefore, we started the simulation after imposing a smooth
Gaussian cut-off
in density which sets in at $\sim 7.3$ kpc.
The latter value corresponds to
the nominal Jacobi tidal radius of the model at the apocenter of
the orbit.

The parameters of the best dwarf galaxy model were chosen after a number of
trial simulations. In those simulations we varied slightly the orbit,
the initial mass and the other halo and disk parameters but tried to keep them as
close as possible to the general trends in cold dark matter
cosmologies, such as those concerning the relation between the virial
mass and concentration of the halo (e.g., Bullock et al. 2001) and the
relations between the different disk parameters (Mo et al. 1998).
In summary, we considered dark halo
virial masses in the range $(0.8-1.6) \times 10^{10} M_\odot$,
concentrations of the dark halo $c=4-30$, different inner slopes of the dark halo $\alpha =0.6-1$,
$\lambda$ parameters $0.04-0.08$,
disk mass fractions $m_{\rm d}=0.01-0.04,$ disk thicknesses $z_{\rm d}/R_{\rm d} = 0.2-0.3$,
pericenter distances $r_{\rm p} = (14-21)$ kpc and
apocenters $r_{\rm a} = (58-60)$ kpc.
We only used some
combinations of these parameters and in adjusting them we relied on the knowledge of their
effect on the evolution gained from the systematic study by
Kazantzidis et al. (2010).

\begin{figure}
\begin{center}
    \leavevmode
    \epsfxsize=8.5cm
    \epsfbox[0 100 350 475]{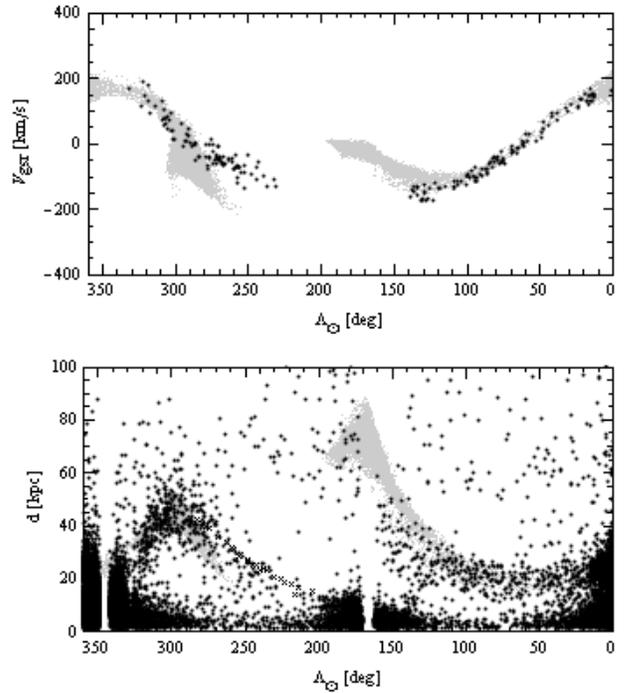}
\end{center}
\caption{Radial velocities ({\it upper panel\/}) and distances from the Sun ({\it lower panel\/})
of the simulated debris ({\it gray points\/})
compared to the data for M giants ({\it black points\/}) from Majewski et al. (2003, 2004) and Law et al. (2005).
The crosses in the lower panel show the distance
measurements for the stars in the leading tail from Belokurov et al. (2006). The $\Lambda_{\odot}$ coordinate
measures the angular distance from Sgr in the direction of the trailing tail (see Majewski et al. 2003).}
\label{debris}
\end{figure}

\begin{figure}
\begin{center}
    \leavevmode
    \epsfxsize=8.cm
    \epsfbox[50 0 260 350]{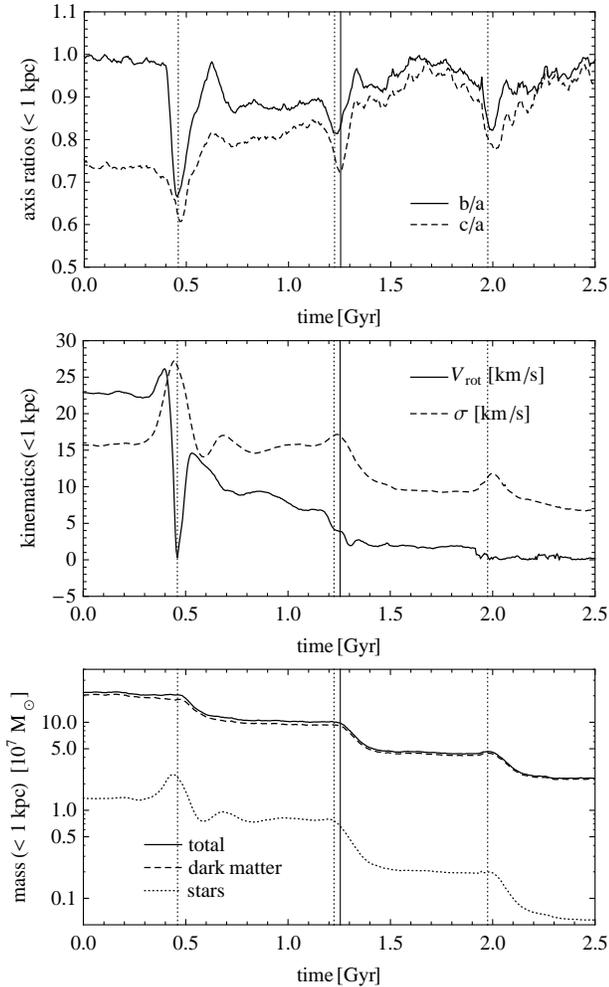}
\end{center}
\caption{Evolution of the simulated dwarf. The panels from top to bottom show the following quantities as a function
of time: the axis ratios, kinematical properties (the rotation velocity around the shortest axis $V_{\rm rot}$ and
1D velocity dispersion $\sigma$), and mass. All quantities were measured on
particles within a distance $r<1$ kpc from the center of the dwarf. Except for the mass, all measurements refer to
the stellar component of the dwarf. The vertical solid lines indicate the presumed
present state of Sgr while the dotted lines mark the pericenter passages.}
\label{evolution}
\end{figure}

\begin{figure*}
\begin{center}
    \leavevmode
    \epsfxsize=15cm
    \epsfbox[0 0 540 550]{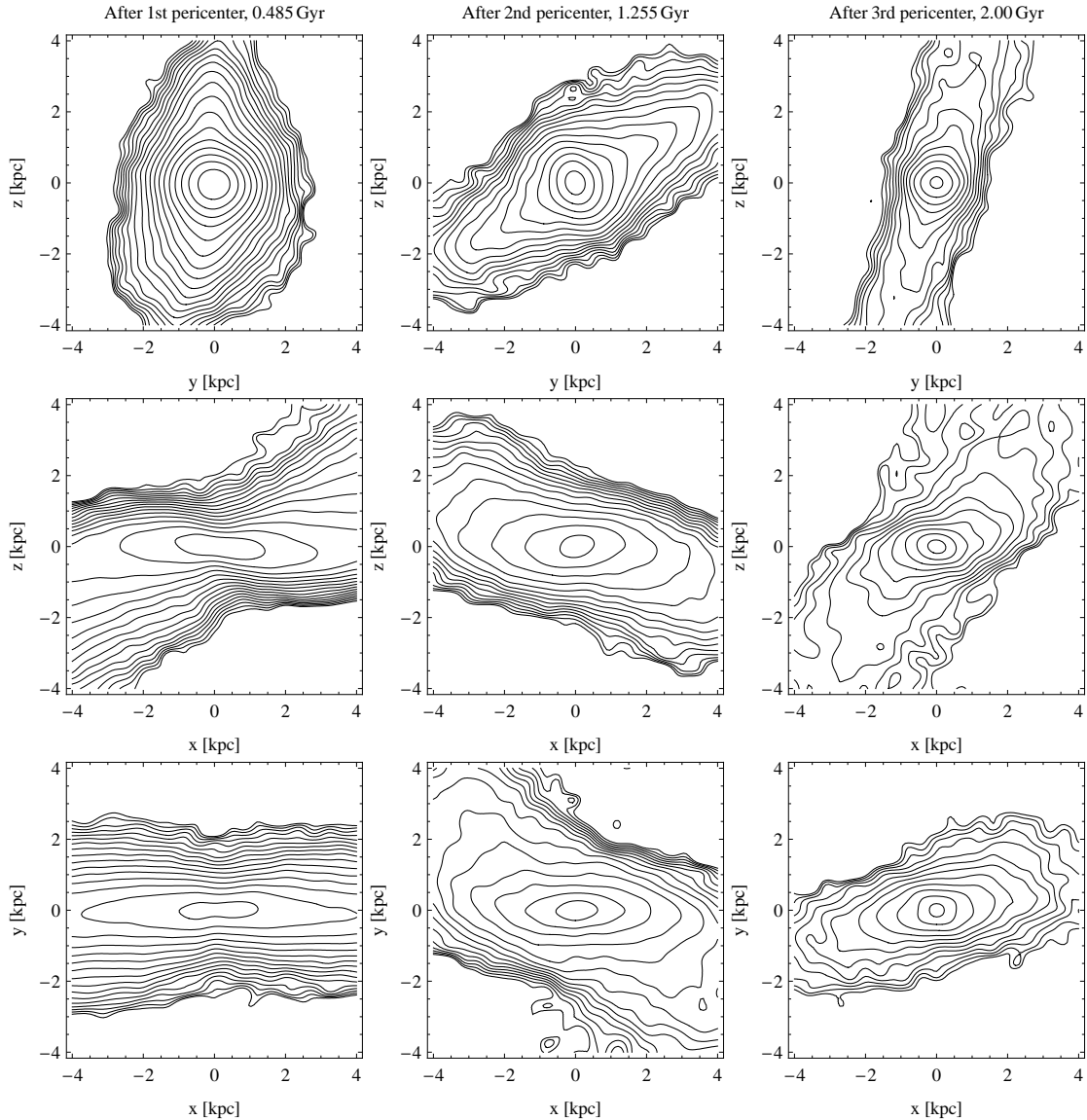}
\end{center}
\caption{Surface density of stars in the simulated dwarf as seen by an observer located at infinity along its longest,
intermediate and the shortest axis (from the upper to the lower row) after the first, second and third pericenter
passage (from the left to the right column).}
\label{surdenall}
\end{figure*}

\begin{figure}
\begin{center}
    \leavevmode
    \epsfxsize=6.8cm
    \epsfbox[0 0 200 600]{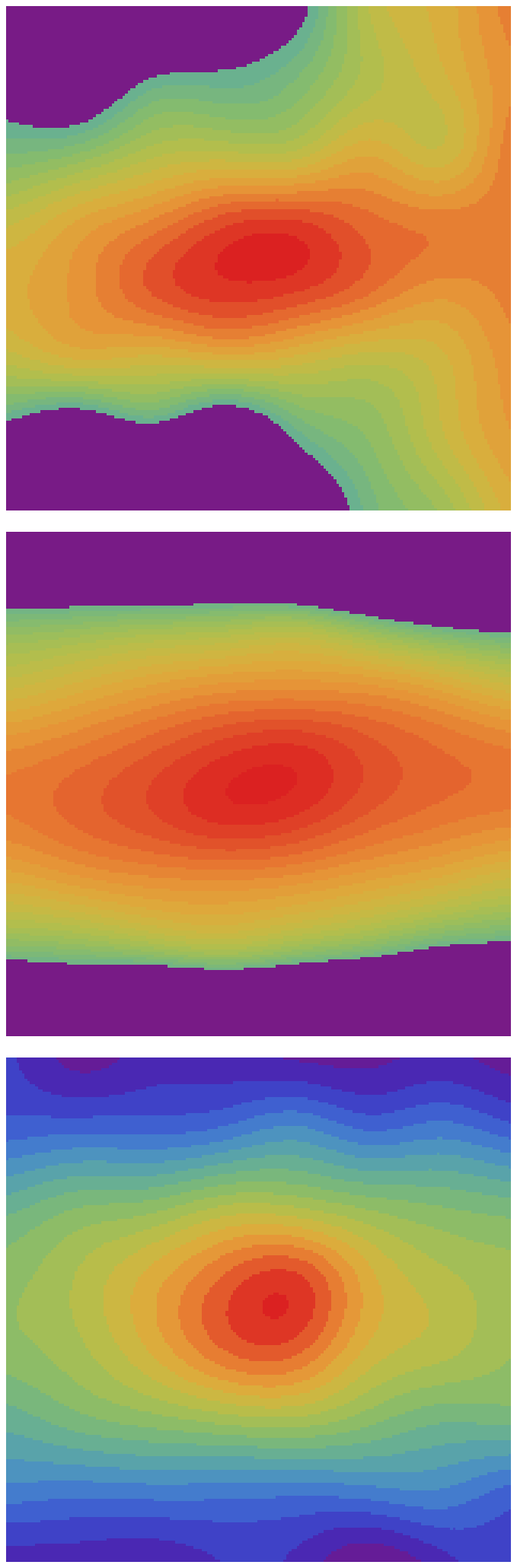}
\end{center}
\caption{({\it Upper panel\/}) The surface density of M giant stars as seen by an observer at the Sun in the real Sgr
(from Figure~4 of Majewski et al. 2003). The flaring
overdensity of stars toward the
right part of the image is due to contamination from Milky Way disk M giants.
({\it Middle panel\/})
The surface density of the simulated dwarf at the present time, 1.255 Gyr from the start of the simulation, right after
the second pericenter.
({\it Lower panel\/}) The surface density of the dwarf's dark matter halo which was initially spherical.
The horizontal direction is the projection of the Sgr orbital plane. The image size in all panels is 14$^\circ$.}
\label{realsim}
\end{figure}

\section{Evolution of the simulated dwarf}

Figure~\ref{overview} shows the simulation output projected onto the orbital plane after 1.255 Gyr of evolution when the
dwarf galaxy position corresponds to the present position of Sgr. The Sun was placed at a distance of 8 kpc
from the Galactic Center (GC) and the simulated Sgr is then at 23 kpc from the Sun (within the range 22-28 kpc of
current best estimates, see e.g., Table 2 in Kunder \& Chaboyer 2009).  The distances of the Sgr center from the GC and
from the Sun are plotted in the upper panel of Figure~\ref{distvel} as a function of time. Note that our orbit is
decaying (the apocenter and pericenter distances decrease slightly) due to dynamical friction,
because we use a live Galaxy model.
The total velocity and the radial velocities of the dwarf measured in the Galactic Standard of Rest frame at the Sun
and at the GC are shown in the lower panel of Figure~\ref{distvel}.

To check if our model is consistent with the available data on the leading and trailing Sgr debris, in
Figure~\ref{debris} we compare the radial velocities (upper panel) and distances (lower panel) of the stars in the
leading and trailing tails to the data. We can see that the simulated trailing tail matches the observed velocities
very well, while in the leading tail the velocities are slightly too large. A similar discrepancy is seen in the
distances of the tails. This behavior is in agreement with the expectations for the spherical Milky Way halo model
assumed here, which cannot account perfectly for all properties of the tails (Law et al. 2005).
However, our purpose here was not to reproduce the tails exactly, but rather to demonstrate that at the time
we chose as corresponding to the present phase of Sgr evolution, the dwarf has already produced most of the debris
required to account for the observed extent of the trailing and leading arm.

Figure~\ref{evolution} illustrates the evolution of the dwarf in time up to 2.5 Gyr from the start of the simulation.
All quantities shown in this Figure were estimated using stars (or dark matter particles) within a fixed radius of 1
kpc from the center of the dwarf.  By adopting this rather low value we make sure that only particles inside the
Sgr core are included at all times ---  i.e., we measure the intrinsic properties of the dwarf
avoiding the contamination by tidally stripped material (the adopted scale of 1 kpc
is dictated by late evolutionary stages when the dwarf is heavily stripped and
therefore much reduced in size; see Figure~\ref{surdenall}).

The most interesting and relevant parameter
for us here is the evolution of the shape of the satellite, which we quantify by the axis ratios $b/a$ and
$c/a$ where $a$, $b$ and $c$ are the longest, intermediate and the shortest axis of the distribution of the stars. To
estimate them, for all simulation outputs (saved every 0.005 Gyr) we determine the directions of the principal axes of
the stellar component using the moments of the inertia tensor. The evolution of the axis ratios is shown in the top
panel of Figure~\ref{evolution}.

Other interesting aspects
of the evolution of the dwarf are related to its kinematics. To measure the kinematics we
introduce a spherical coordinate system ($r$, $\theta$, $\phi$) centered on the dwarf
so that the $z$-axis is along the shortest axis of the
stellar distribution and the angle $\phi$ is measured in the $xy$ plane.  We then calculate the rotation velocity
around the shortest axis $V_{\rm rot}=V_{\phi}$ and the dispersions
$\sigma_r$, $\sigma_{\theta}$ and $\sigma_{\phi}$ around the
mean values. We combine the dispersions into the 1D dispersion parameter $\sigma = [(\sigma_r^2 + \sigma_{\theta}^2 +
\sigma_{\phi}^2)/3]^{1/2}$, which measures the amount of random motion in the stars. The evolution of these two
quantities is illustrated in the middle panel of Figure~\ref{evolution}.

The bottom panel of the figure shows the evolution of the stellar, dark and total mass of the dwarf contained inside a
radius of 1 kpc. The strongest decrease in all of the masses occurs around pericenters, i.e., after 0.46, 1.225 and 1.975
Gyr from the start of the simulation (vertical dotted lines in Figure~\ref{evolution}). We note that the mass loss
in stars follows that in dark matter.

The evolution of the shape is further illustrated in Figure~\ref{surdenall}. Here we plot the surface density
distribution of the stars selected within $r<6$ kpc
from the center of the dwarf as seen along the longest ($x$), intermediate ($y$) and shortest ($z$) axis
(from the top to the bottom row) of the stellar distribution. In columns we show the results right after
the first, second and third pericenter.
The second column corresponds to the simulation output shown in
Figure~\ref{overview} which we choose as best
corresponding to the present time.

The analysis of the evolution of different quantities in Figure~\ref{evolution} and the surface density plots in
Figure~\ref{surdenall} reveals a clear picture of the fate of the dwarf. The shape of the dwarf galaxy disk evolves so
that after the first pericenter a bar is formed ($b/a=c/a$). This bar soon transforms into a triaxial, but still
prolate shape. At the second pericenter the bar becomes stronger again and at the moment shown in
Figure~\ref{overview}, which corresponds to the present state of Sgr,
the dwarf is still quite elongated with $b/a=0.83$ and $c/a=0.72$ inside the inner 1 kpc. Note that
these values depend on radius and would be $b/a=0.64$ and $c/a=0.57$ when measured within the disk initial
scalelength $r < R_{\rm d} = 2.3$ kpc. Right after
the second pericenter the stellar component becomes almost spherical with $b/a$ and $c/a$ both
very close to unity. The shape becomes a little elongated again at the third pericenter because the tidal forces are at
their maximum.

The changes of the shape of the stellar component are accompanied by a strong mass loss and decreasing rotation velocity
$V_{\rm rot}$ of the stars. The velocity dispersion also decreases with time (except for peaks at
pericenters due to tidal heating) because of mass loss. However, the ratio $V_{\rm rot}/\sigma$ drops below unity already
at the first pericenter and remains like that until the end. At the time we chose as corresponding to the present
evolutionary stage of the dwarf, after 1.255 Gyr from the start of the simulation, the remnant rotation in
the core ($r < 1$ kpc) is very low, of the order of 4 km s$^{-1}$ with $V_{\rm rot}/\sigma = 0.22$ (0.4 at the scale
of $R_{\rm d} = 2.3$ kpc). Note that at the first pericenter, when the bar
forms, the rotation velocity drops to zero. Right after this pericenter the orientation
of the bar is such that there is a strong tidal torque from the Milky Way that speeds up the bar in the same
direction the disk was initially rotating. This phenomenon is similar to the one
discussed by Kazantzidis et al. (2010, see their section 5.2 and figure 13) in a
similar context.

During the whole evolution the shortest axis of the stellar component of the dwarf
is almost perpendicular to the orbital plane
because the initial disk was inclined by only 10$^\circ$ to the orbital plane. Since the bar forms in the plane of the disk
the longest axis lies almost in the orbital plane. Because the tidal tails also lie in this plane we see them clearly in
the bottom row of Figure~\ref{surdenall}. The bar happens to be oriented almost perpendicular to the line of sight of
the observer at the Sun (see Figure~\ref{overview})
at the present time so such an observer sees Sgr almost along the
intermediate axis (the latter as shown in the middle row of Figure~\ref{surdenall}).
This observer's view of the simulated dwarf is shown in the middle panel of Figure~\ref{realsim}.
For comparison, in the upper panel we plot the surface density of the real Sgr M giants from Majewski et al. (2003). The
inner isodensity
shapes are very similar: strongly elongated, both with ellipticity $e=1-b/a$ of the order of 0.6, where $b$ and $a$
are now measured along the major and minor axis of the image from the projected surface density contours at the radius of
5$^\circ$.
Note that the outer isodensity shapes for the actually observed Sgr
are perturbed by contamination of stars from the Milky Way disk.

The coordinate system chosen for these images measures the angles along the $\Lambda_{\odot}$ and
$B_{\odot}$ coordinates associated with the orbital plane of Sgr projected on the sky (Majewski et al. 2003).
Thus the orbital plane lies in the horizontal direction of the plots and
both images show slight inclination with respect to this plane. The inclination of the simulated dwarf measured
to be about 6$^\circ$ at the radius of 5$^\circ$ from the center agrees very well with the ``canting angle"
of the Sgr main body with respect to its orbital plane measured for the real data by Majewski et al. (2003).
The inclination of the simulated image was reproduced
by adopting the initial inclination of the dwarf disk of 10$^\circ$ with respect to the orbital plane
(by rotating the disk around the $X$ axis of Figure~\ref{overview} so that the half of the disk closer to
the Galactic center is above the plane of Figure~\ref{overview}; see also Figure~\ref{initial}).
The angle seen at the present time is
expected to be lower than the initially adopted value because in the observed image the intrinsic
surface density contours of the core are affected by the tidally distorted distribution of stars at larger projected
radii. These tidal extensions are obviously along the orbit and therefore decrease the observed inclination.
Note, however, that the initial inclination angle cannot be much higher than the 10$^\circ$ we adopt
since then the core would be more inclined and would form a pronounced S-shape with the tails that is not seen in the
data.

In summary, since the formation of the bar can only occur in disks, if one starts
with a spherical dwarf model the
observed elongated shape of Sgr is
very difficult to reproduce. To further support this statement, in the lower
panel of Figure~\ref{realsim} we show the surface density distribution of dark matter as it would be seen at present by
an observer at the Sun. The dark matter halo of the dwarf was initially spherical and, clearly, the surface density
contours remain spherical in the center at present and are very different in shape from the contours of the
distribution of the stars. This strongly suggests that the stellar component of Sgr must have been very different from
spherical when the dwarf entered its orbit around the Galaxy.  In addition, right after the second pericenter the
stellar component quickly transforms into a much more spherical shape; thus, according to this model,
Sgr could not have been a satellite for much more than 1.5 orbital times in its {\it present\/} orbit,
in which it must have just passed its second pericenter around the Milky Way.

\begin{figure}
\begin{center}
    \leavevmode
    \epsfxsize=8.5cm
    \epsfbox[0 0 290 290]{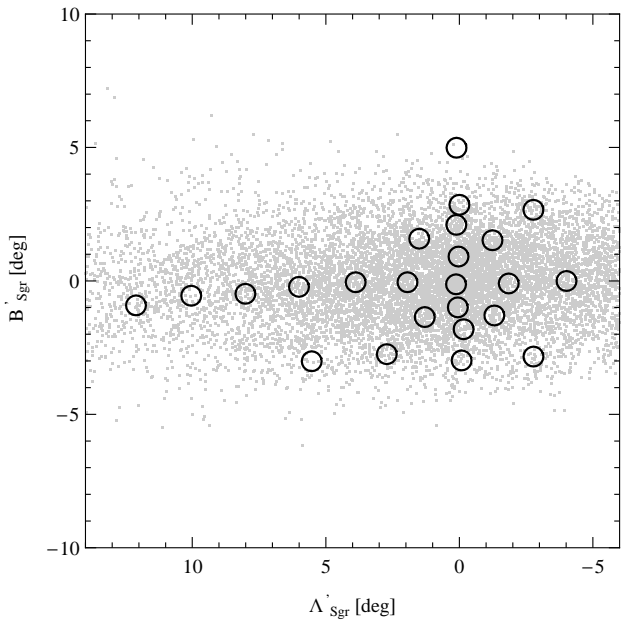}
\end{center}
\caption{View of the simulated Sgr core ({\it gray dots\/}) and
the positions of the fields analogous to those in the real observations of Sgr kinematics
by Frinchaboy et al. (2010; {\it black circles\/}). The coordinate system is that of
$\Lambda'_{\rm Sgr}$ and $B'_{\rm Sgr}$, where $\Lambda'_{\rm Sgr}$ is measured along the
major axis of the Sgr image, i.e. the positions of the stars were rotated clockwise by the angle of 6$^\circ$
with respect to those used in Figure~\ref{realsim}.}
\label{positions}
\end{figure}

\begin{table*}
\caption{Kinematics of the simulated Sgr}
\label{kinsim}
\begin{center}
\begin{tabular}{lrrrrr}
\hline
\hline
Field & $N_{\rm tot}$ & $N_{\rm sel}$ & $R$ [deg] & $\langle V_{\rm gsr} \rangle$ [km s$^{-1}$] &
$\sigma_{V_{\rm gsr}}$ [km s$^{-1}$]  \\
\hline
Major+00   &  1577 & 102 & 0.29 & $ 171.81  \pm 1.94 $ &  $ 19.64  \pm  1.38 $  \\
Major$-$04 &   608 & 110 & 3.98 & $ 172.06  \pm 1.49 $ &  $ 15.66  \pm  1.06 $  \\
Major$-$02 &   930 &  86 & 1.81 & $ 170.23  \pm 1.87 $ &  $ 17.33  \pm  1.33 $  \\
Major+02   &   931 & 182 & 1.98 & $ 172.79  \pm 1.45 $ &  $ 19.60  \pm  1.03 $  \\
Major+04   &   680 &  82 & 3.91 & $ 170.03  \pm 1.65 $ &  $ 14.91  \pm  1.17 $  \\
Major+06   &   433 &  57 & 6.03 & $ 166.21  \pm 2.07 $ &  $ 15.62  \pm  1.48 $  \\
Major+08   &   249 &  45 & 8.00 & $ 160.51  \pm 2.32 $ &  $ 15.55  \pm  1.66 $  \\
Major+10   &   159 &  24 & 10.01& $ 154.61  \pm 3.14 $ &  $ 15.39  \pm  2.27 $  \\
Major+12   &   101 &  14 & 12.08& $ 152.94  \pm 4.58 $ &  $ 17.15  \pm  3.36 $  \\
Minor$-$03 &   208 &  31 & 2.91 & $ 167.79  \pm 3.22 $ &  $ 17.95  \pm  2.32 $  \\
Minor$-$02 &   533 &  56 & 1.79 & $ 166.05  \pm 2.32 $ &  $ 17.36  \pm  1.65 $  \\
Minor$-$01 &  1008 &  89 & 0.97 & $ 167.70  \pm 2.06 $ &  $ 19.47  \pm  1.47 $  \\
Minor+01   &  1112 &  56 & 0.93 & $ 175.55  \pm 2.61 $ &  $ 19.55  \pm  1.86 $  \\
Minor+02   &   405 &  48 & 2.05 & $ 174.33  \pm 2.25 $ &  $ 15.62  \pm  1.61 $  \\
Minor+03   &   224 &  49 & 2.82 & $ 172.41  \pm 2.53 $ &  $ 17.70  \pm  1.81 $  \\
Minor+05   &     6 &   6 & 4.85 & $ 167.49  \pm 3.15 $ &  $ 7.71   \pm  2.44 $  \\
NW+02      &   671 &  62 & 1.96 & $ 167.62  \pm 2.02 $ &  $ 15.90  \pm  1.44 $  \\
SW+02      &   669 &  60 & 1.83 & $ 167.69  \pm 2.21 $ &  $ 17.09  \pm  1.57 $  \\
SE+02      &   724 &  88 & 1.88 & $ 174.66  \pm 1.53 $ &  $ 14.38  \pm  1.09 $  \\
NE+02      &   550 &  37 & 2.18 & $ 179.11  \pm 2.62 $ &  $ 15.95  \pm  1.88 $  \\
NW+04      &   195 &  17 & 3.86 & $ 169.48  \pm 3.47 $ &  $ 14.30  \pm  2.53 $  \\
SW+04      &   175 &  12 & 3.96 & $ 155.50  \pm 4.40 $ &  $ 15.24  \pm  3.25 $  \\
SE+04      &   193 &  34 & 3.86 & $ 170.95  \pm 2.49 $ &  $ 14.49  \pm  1.78 $  \\
ESE+07     &    67 &  25 & 6.28 & $ 168.27  \pm 2.58 $ &  $ 12.88  \pm  1.86 $  \\
\hline
\end{tabular}
\end{center}
\end{table*}

\section{Modeling of the kinematics}

Recently Frinchaboy et al. (2010) have measured the radial velocities and velocity dispersions of stars in a number of
pointings in the field of the Sgr main body. We model these data later in this section.  First, we produce similar mock
kinematical data sets by probing the simulated dwarf in the same way as Frinchaboy et al.
Figure~\ref{positions} shows the positions of a fraction of
stars in the simulated dwarf as they would be observed from the Sun. The
coordinate system adopted here was such that the $x$ axis is along the major axis of the image of the dwarf's stellar
component and the $y$ axis is along the minor axis (in analogy
with the $\Lambda'_{\rm Sgr}$ and $B'_{\rm Sgr}$ coordinates
adopted in the real observations by Frinchaboy et al.). The black circles indicate the model-sampled fields,
which are analogous to those applied in the real Frinchaboy et al. observations and are
listed in the first column of Table~\ref{kinsim} with the same names as in that reference.
Because there are no multiple tidal streams in our simulation,
in this direction of the sky only the main body of Sgr is seen and there is almost
no contamination from tidal debris along the line of sight (except for the stars presently being stripped).
By using only the stellar particles of the simulated dwarf we also avoid the contamination from the Milky Way.
Thus there is no need to introduce any cut-off in velocity for the model kinematical measurements, as was done
in Frinchaboy et al.

The total numbers of stars selected within the fields, $N_{\rm tot}$, are listed in the second column of
Table~\ref{kinsim}. Only in the case of the field Minor+05 is the number of stars
smaller than observed (6 versus 9). This turns out to be understandable in the light of the fact that this field
is likely the one most contaminated by Milky Way stars (see below and Frinchaboy et al. 2010).
In all other fields the number
of stars in the simulated Sgr is much larger than in the actually observed samples. To create mock data sets as
similar as possible to the real ones for all fields (except Minor+05 where we take all 6 stars available)
we randomly select a number of stars exactly the same as in the real data. The numbers of selected stars $N_{\rm sel}$
are listed in the third column of Table~\ref{kinsim}. The next columns of the Table list the mean projected
distance of the stars in a given field from the center of the dwarf and the mean velocity and velocity
dispersion obtained in the field with their respective standard errors.

\begin{figure}
\begin{center}
    \leavevmode
    \epsfxsize=8.5cm
    \epsfbox[30 10 260 275]{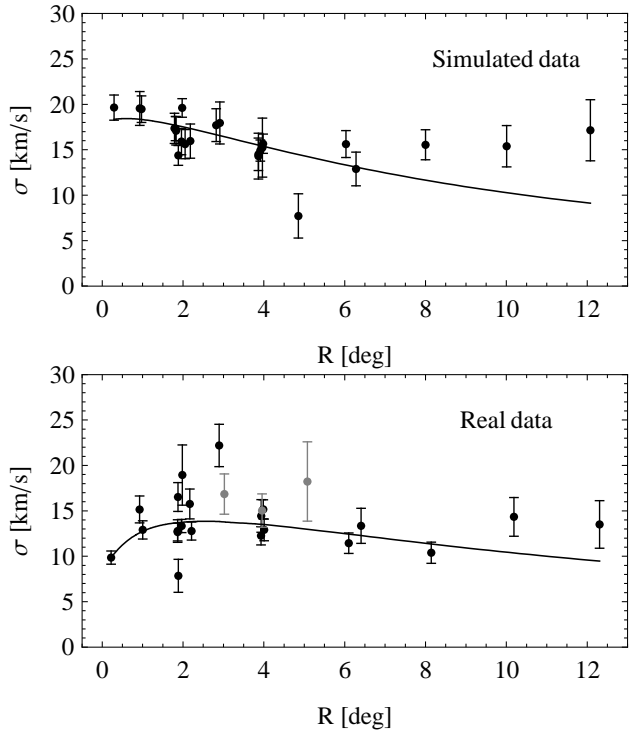}
\end{center}
\caption{The line-of-sight velocity dispersion of the simulated dwarf (upper panel) and the real
Sgr (lower panel) as a function of radial distance from the center $R$. Solid lines show the best-fitting solutions of
the Jeans equation. The data points marked in gray in the lower panel are most contaminated by the Milky Way stars
and were excluded from the fit.}
\label{dispersions}
\end{figure}

\begin{figure}
\begin{center}
    \leavevmode
    \epsfxsize=8.5cm
    \epsfbox[0 0 230 230]{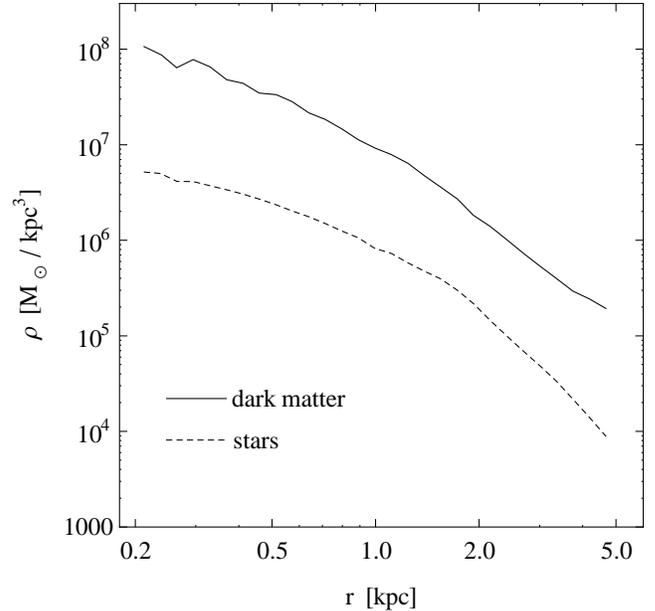}
\end{center}
\caption{The density profiles of stars ({\it dashed line\/}) and dark matter ({\it solid line\/})
in the simulated dwarf up to $r=5$
kpc, which corresponds to the angular scale of 12$^\circ$ probed by the Frinchaboy et al. (2010) kinematical measurements.}
\label{densityprofiles}
\end{figure}

The velocity dispersion profile of the simulated dwarf obtained in this way is shown in the upper panel of
Figure~\ref{dispersions}. We model the profile using solutions of the Jeans equation (see, e.g., {\L}okas 2002;
{\L}okas et al. 2005; {\L}okas 2009), assuming that mass follows light and that the anisotropy parameter
$\beta$ is constant
with radius. The anisotropy parameter measures the amount of radial versus circular orbits in the stellar population
and we define it in the standard way as $\beta = 1 - (\sigma_\theta^2 + \sigma_\phi^2)^2/(2 \sigma_r^2)$. By adjusting
the solutions to the measured profile we estimate the total mass $M$ and the anisotropy parameter $\beta$. The
mass-follows-light assumption is well justified by the measurements of the density profiles of stars and dark matter
within 5 kpc (corresponding to the angular scale of 12$^\circ$ in the data). The density profiles are plotted in
Figure~\ref{densityprofiles}, where one can see that they follow each other. We have also verified that the anisotropy
parameter does not strongly vary with radius and is contained within $0 < \beta < 0.4$ for radii 0 kpc $ < r < 5$ kpc
with the average value $\beta = 0.17 \pm 0.13$.

\begin{figure}
\begin{center}
    \leavevmode
    \epsfxsize=8.5cm
    \epsfbox[0 0 230 230]{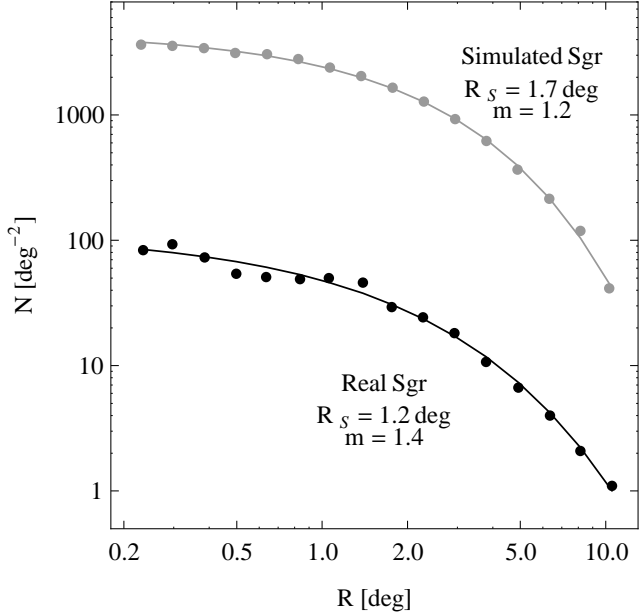}
\end{center}
\caption{Surface density profiles of the simulated dwarf ({\it gray points\/})
and the real Sgr ({\it black points\/}) measured within circular projected radii.
Solid lines show the best-fitting S\'ersic profiles.}
\label{surdenprof}
\end{figure}

To perform this analysis we adopt the density profile of the stars as obtained by deprojection of
their surface number density profile. This profile measured from the simulation is shown as gray points in
Figure~\ref{surdenprof}. The measurements were done up to projected $R<12^\circ$, the distance of the furthest kinematical
measurement along the major axis by Frinchaboy et al. (2010). Interestingly, the profile does not yet flatten at
these largest distances; such flattening would signify the transition to tidal streams.
The model profile is well fitted by the S\'ersic (1968) formula $N(R) = N_0
\exp[-(R/R_S)^{1/m}]$ with the S\'ersic radius $R_{\rm S} = 1.7^\circ$ and shape parameter $m = 1.2$.

The best-fitting solution of the Jeans equation is shown in the upper panel of Figure~\ref{dispersions} as the solid line.
The best-fitting parameters are $M = (8.1 \pm 0.5) \times 10^8 M_{\odot}$ and $\beta = 0.12^{+0.12}_{-0.16}$. Note
that the quality of the fit is quite poor, with $\chi^2/N=46/22$. The $1\sigma$ errors were estimated from $\Delta
\chi^2$ statistics.  The anisotropy estimate agrees very well (within errors) with the averaged value
$\beta = 0.17$ measured from the simulation. The mass within $R < 12^\circ$ (which corresponds to 5 kpc) of this
best-fitting model is $M = 7.3 \times 10^8 M_{\odot}$, significantly higher than the mass actually contained within
this radius in the simulated dwarf, $M = 4.1 \times 10^8 M_{\odot}$.  Note that for objects observed perpendicular to
the bar we actually expect to underestimate the mass from the Jeans analysis rather than overestimate it ({\L}okas et
al. 2010). This is not the case for the present state of Sgr since it is close to pericenter and therefore departs strongly
from equilibrium. These departures manifest themselves in the strong mean radial velocity signal (in the spherical
coordinates related to the dwarf, introduced in section 3, i.e.
the dwarf is
expanding due to the action of strong tidal forces) and the increasing velocity dispersion at the outer parts, which are
affected by tidally stripped material.

\begin{figure}
\begin{center}
    \leavevmode
    \epsfxsize=8.5cm
    \epsfbox[0 10 290 290]{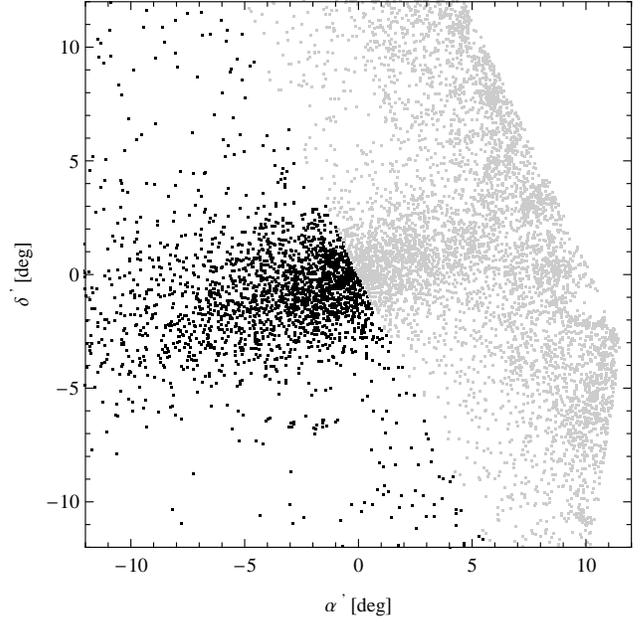}
\end{center}
\caption{Positions of the stars from the 2MASS survey of Sgr. The coordinates $\alpha'$ and $\delta'$ are measured
along the RA and Dec with respect to the Sgr's center. Black dots indicate the stars used for the calculation of
the surface number density profile shown in Figure~\ref{surdenprof}. Stars marked in gray were rejected due to the
contamination by the Milky Way disk.}
\label{selectedstars}
\end{figure}

We proceed in an analogous way to fit the velocity dispersion profile obtained from the real data for Sgr,
shown in the lower panel of Figure~\ref{dispersions}. In gray we marked the data points corresponding to the fields
Minor$-$03, Minor+05 and NW+04 where the contamination by Milky Way stars exceeds 50\% according to
the estimates by Frinchaboy et al. (2010).
The surface density profile of the stars measured from the M
giants (Majewski et al. 2003) is shown in Figure~\ref{surdenprof} as black dots.  To obtain an estimate of the
profile out to circular radii of $R < 12^\circ$ and yet avoid the contamination from the Milky Way disk
(seen as the flaring of the outer contours in the right side of Figure~\ref{realsim}, top panel) we measured the density
of stars only on one side (that at higher Galactic latitude)
of the dwarf, as illustrated in Figure~\ref{selectedstars}: only the stars shown in black were
used for the measurements. The rejected stars, shown in gray, were separated by a line parallel to the Milky Way disk.
A S\'ersic profile has been fitted to the data and the best-fitting parameters are: $R_{\rm S} = 1.2^\circ$ and $m = 1.4$.
Note that in spite of the different parameters (the S\'ersic profile is strongly degenerate for $R_{\rm S}$ and $m$) the
shapes of the density profiles of the stars in the simulated and real Sgr are very similar.

Assuming that mass follows light we fit the velocity dispersion profile of Sgr by adjusting the total mass and
anisotropy. When performing this fit we have excluded the data points marked in gray in the lower panel of
Figure~\ref{dispersions}. The best-fitting parameters are $M = (6.1 \pm 0.5) \times 10^8 M_{\odot}$ and $\beta =
-0.68^{+0.24}_{-0.32}$. The corresponding best-fitting
dispersion profile is shown with a solid line in the lower panel of Figure~\ref{dispersions}.
The mass within $R < 12^\circ$ of this best-fitting model is $M = 5.2 \times 10^8 M_{\odot}$,
similar to the one of the simulated dwarf. The quality of the fit is also poor this time, with $\chi^2/N =
49/19$. This is due to the fact that the dispersion data points are very scattered around the mean fit,
which may be due to some contamination from the Milky Way stars (still likely present in some fields,
as shown by Frinchaboy et al.,
in spite of the $3\sigma$ clipping applied to the velocity
measurements by those authors and our rejection of most contaminated data points)
and the tidally stripped stars, as mentioned above.
The total mass we find is about 40\% lower than the early rough estimate of Ibata et al. (1997) who found the
mass to be of the order of $10^9 M_{\odot}$.
Our estimate is however very close to the more precise value $M = (5.8 \pm 0.5) \times 10^8 M_{\odot}$ obtained
by Majewski et al. (2003) from a single velocity dispersion measurement by following the Richstone \& Tremaine (1986)
version of the King (1966) formalism. Our value agrees also very well with the one recently obtained by Law \& Majewski
(2010) from the dispersion of velocities in the tidal tails. Within 4$^\circ$ they find $M = 2.5^{+1.3}_{-1.0}
\times 10^8 M_{\odot}$ while our mass profile gives $2.1 \times 10^8 M_{\odot}$.

Figure~\ref{rotation} compares the rotation curves of the simulated and real Sgr. The points with error bars are the
measurements of the mean velocity of the stars along the major axis in the real Sgr from
Frinchaboy et al. (2010). The solid line connects analogous
values measured in the simulation and listed in Table~\ref{kinsim}, but without the errors. The match is very good.
Note that at the stage of the evolution selected as corresponding to the present state of Sgr, the stellar
component of the
simulated dwarf had very little remnant rotation, of the order of 4 km s$^{-1}$ (see the second panel of
Figure~\ref{evolution}). This intrinsic rotation is very much obscured in Figure~\ref{rotation} by the velocity
gradient due to tidal tails pointed in the opposite direction. This reversal of the intrinsic rotation is similar to
the effect seen in Leo I ({\L}okas et al. 2008). Note that the observed velocity trend shown in Figure~\ref{rotation}
could not be reproduced by the model proposed by Pe\~narrubia et al. (2010) where the intrinsic rotation is still very
high (of the order of 20 km s$^{-1}$). We have also verified that the observed velocity trend
cannot be reproduced by a similar model as presented here but with the disk spinning in a retrograde manner.
In such a case the stars stripped
from the dwarf form tidal tails of similar length but with a much wider distribution of stars in distance and
velocity than observed. In addition, a disk on a retrograde orbit does not form a bar and it is more difficult to
remove rotation from it. Such significant remnant rotation is then clearly seen in the velocity trend
right from the center of Sgr and has the same sign as the tidally induced velocity shear
seen in Figure~\ref{rotation}.

\section{Summary and discussion}

We presented a plausible model for the
inner structure and kinematics of the Sgr dwarf galaxy. By adopting an
initial stellar distribution in the form of a disk rather than a
spheroid, this model reproduces for the first time the present, very
elongated shape of the Sgr stellar core. The satellite orbit around
the Milky Way was chosen so that its present distance and velocity
with respect to the Sun agree well with the observed values. Our model also has a
total mass similar to that estimated from the velocity dispersion
profile of the real Sgr as well as a velocity gradient along the major
axis that matches that seen in the real data.  By requiring the dwarf
to produce enough tidally stripped debris to match most of the
observed extent of the Sgr stream and at the same time still preserve
its highly non-spherical shape we put constraints on the number of
pericenters Sgr could have passed until the present
in its current orbit. For the present
model to work, Sgr must have had only two perigalacticon passages.
Indeed, if there had only been one pass not enough tidal debris would
be produced, whereas after three passages the Sgr core would have
already become significantly more spherical and thus inconsistent with
observations.

Let us note that although the parameter space we explored was quite wide, we
cannot absolutely exclude the possibility that the Sgr dwarf entered its orbit
around the Milky Way with a spherical stellar component. The numerical
experiments we performed strongly suggest, however, that it is very difficult
to reproduce its present, elongated shape through the action of tidal forces on
a spherical stellar distribution. For this to occur, the tidal forces would have
to distort the dwarf down to the very center implying that is was weakly self-bound.
Whenever we tried to significantly lower the self-gravity of the dwarf, e.g. by
adopting a dark matter halo of low concentration, we found it to be very quickly
destroyed by tidal forces; such an object would not survive long enough on its
present orbit to produce enough tidal debris.

\begin{figure}
\begin{center}
    \leavevmode
    \epsfxsize=8.5cm
    \epsfbox[0 10 285 165]{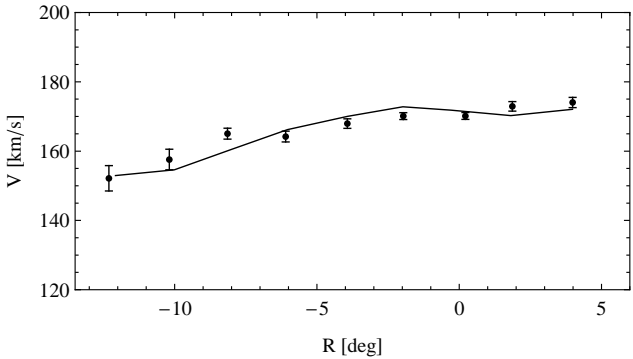}
\end{center}
\caption{The rotation curve measured along the major axis of the Sgr image. The points with error bars show the
kinematic measurements for real Sgr from Frinchaboy et al. (2010). The solid line shows the measurements in the simulated
dwarf performed in the same way as in the real observations, as shown in Figure~\ref{positions}.}
\label{rotation}
\end{figure}

Our model also reproduces in a satisfactory way the density profile of
the stars, the total luminosity and the mass-to-light ratio
of the Sgr core. For the
real Sgr in its present state, using our total mass estimate and adopting
the luminosity of $2 \times 10^7 L_{\odot}$ (Mateo et al. 1998;
Majewski et al. 2003), we get $M/L$ of the order of 30 in solar units.
As in the case of the mass, this is significantly
lower than the estimate of $M/L$ of 50 by Ibata et al. (1997) but agrees well with the estimate by
Majewski et al. (2003) based on core-fitting.
To find the corresponding value for
the simulated dwarf we need to translate the mass of stars into luminosity. Since the stellar $M/L$ are poorly known,
especially for systems containing many stellar populations of different age, we can conservatively adopt $M/L$ in the
range of $(1-3) M_{\odot}/L_{\odot}$ for the stars. The mass of stars within 5 kpc is $3.2 \times 10^7 M_{\odot}$ so the
luminosity is in the range $(1.1-3.2) \times 10^7 L_{\odot}$ in good agreement with the true present luminosity of Sgr.
The mass-to-light ratio is then in the range $(13-39) M_{\odot}/L_{\odot}$, which includes our value obtained for
the real Sgr.

The present shape of Sgr must be bar-like with the bar major axis almost perpendicular to our line of sight.
If it were a disk seen almost edge on, its rotation curve would be very different: we would see a strong rotation signal
at the very center of the dwarf which could become dominated by the velocity gradient of tidal origin only at the outer
parts. What is observed is the rotation velocity increasing very slowly with distance along the major axis
(Frinchaboy et al. 2010). This is
almost entirely caused by the tidally induced velocity gradient of stripped debris, which obscures the remnant intrinsic
rotation if any is present. Such a low intrinsic rotation is only possible in a prolate shape, where radial orbits
dominate the motion of the stars. In the case of Sgr seen from the Sun, this motion is almost perpendicular to the line
of sight.

Recently, Pe\~narrubia et al. (2010) proposed a model of Sgr that also initially contained a disk, with the purpose of
addressing the origin of the bifurcation in the leading Sgr arm visible in Sloan Digital Sky Survey data
(e.g., Belokurov et al. 2006; Yanny et al. 2009).
They showed that if their initial disk was inclined by
about 20$^\circ$ with respect to the orbital plane then the bifurcation would be reproduced.
While an intriguing hypothesis for the origin of the enigmatic bifurcation, our
work suggests that the
initial inclination of the disk was probably about a factor of two lower since otherwise the stellar component
forms a pronounced S-shape that is not observed in the real data. In addition, at the stage corresponding to the present
time their disk retains a significant amount of rotation, of the order of 20 km s$^{-1}$ which does
not match the rotation curve measured by Frinchaboy
et al. (2010) along the Sgr major axis and reproduced by our model in Figure~\ref{rotation}.
The Pe\~narrubia et al. (2010) model also does not
include the effect of dynamical friction, which may significantly change
the structure of the leading arm. It is therefore
not obvious whether a disk model that accurately reproduces the core structure
of Sgr might also be able to produce a bifurcation similar
to that observed.

The orbit that we chose for the progenitor of Sgr is not supposed to simulate its initial cosmological
infall into the Milky Way halo. It is instead meant to model the time over which the transformation of the stellar
component has occurred. Constrained cosmological simulations of the Local Group (Klimentowski et al. 2010)
show that the wide majority of satellites surviving
to $z=0$ are on orbits with apocenters exceeding 100 kpc, namely larger than what we have assumed for the orbit of Sgr.
Likewise, most satellites, especially those accreting after $z=1$ (as is implicitly required
for the case of Sgr by the fact that
our model suggests that it should have performed only two pericenter passages in its current orbit by now),
have virial masses an
order of magnitude below the
initial mass of the progenitor of Sgr (Klimentowski et al. 2010).
However, the apparently unusual orbit and
rather high mass are mutually explained once we try to place Sgr in the more general context of satellite accretion.

It is likely that Sgr was indeed accreted on an orbit having an apocenter exceeding 100 kpc, but this orbit was rapidly
eroded by dynamical friction owing to Sgr's large mass, which enabled Sgr to acquire
the apocenter that we have assumed. During this early time
tidal mass loss occurred mostly in the halo,
as shown by $N$-body simulations (e.g., Mayer et al. 2001). Hence, such mass loss
would not change the results of our work
concerning the evolution of the stellar component of the core.  Note however, that some
mass loss in stars and the formation of stellar streams could take place even before the phases of evolution
we study in this paper. Thus, more pericenter passages than two may have occurred in total, but the earlier ones
at large radii will have minimal effect on the internal structure of Sgr (Kazantzidis et al. 2010).

Figure~\ref{distvel} shows that the effect of dynamical friction is moderate for our initial mass of
$1.6 \times 10^{10} M_\odot$, but if the mass of Sgr before infall was only a factor of a few times
larger ($\sim 3-5 \times 10^{10} M_\odot$), dynamical friction could have reduced its apocenter by at least
a factor of 2 after the first actual pericenter (see, e.g., Colpi et al. 1999; Jiang \& Binney 2000;
Taffoni et al. 2003 --- once
the mass ratio between primary and secondary approaches 20:1 or less the dynamical friction timescale becomes
much shorter than the Hubble time). Therefore in this scenario Sgr would have its current,
unusually tight orbit by
the present time exactly because of the rather
large mass of its progenitor, which implies much more dynamical
friction than for typical satellites.

A higher initial virial mass before infall is also strongly suggested by the quantitative relation between
stellar mass and halo mass suggested by statistical methods, such as halo abundance matching using stellar masses
from SDSS (Guo et al. 2009; Sawala et al. 2010), as well as by structural analysis of individual dwarfs in the Local
Group and, more generally, in the nearby Universe (Mayer \& Moore 2004; Oh et al. 2008; McGaugh \& Wolf 2010).
Indeed, the first of the cited methods implies that typical stellar masses are $< 0.05-1$\% of the halo mass
for halos with virial masses close to $10^{10} M_\odot$, while the second line of evidence,
which has less
statistical power but does not suffer from incompleteness problems due to the low surface brightness of typical
dwarfs, would suggest $\sim 1-2\%$ of the halo mass to be typical (with significant scatter present);
this places
our choice of the initial condition $m_{\rm d}=0.02$ towards the high end of allowed values. For comparison,
detailed mass modeling of two of the best studied, isolated dIrr galaxies in the Local Group, NGC 6822 and NGC 3109,
with circular velocities close to that assumed for the progenitor of Sgr ($40-60$ km s$^{-1}$), would also
yield $m_{\rm d} \sim 0.01$ (Valenzuela et al. 2007). This marginal discrepancy is easily resolved if one assumes that
the initial halo mass before infall was at least a factor of 2 larger than what we have assumed here, thus
placing Sgr among the typical disky dwarfs in terms of the initial stellar to halo mass ratio.

Although our disk mass fraction $m_{\rm d} =0.02$ may appear high in terms of stellar mass, the whole baryonic mass
of dIrr galaxies (including gas) is a factor of a few higher. Since our simulation is collisionless we could not
model explicitly the gas component and our intermediate value reflects the baryonic disk mass after it has been
affected by the tidal field.
There are convincing arguments, both observational and theoretical, that not all of
the gas initially present in the dwarf was converted into stars; some of it must have
been lost soon after the dwarf had become a satellite of the Milky Way. Simulations including hydrodynamics (e.g., Mayer
et al. 2006, 2007) suggest that this gas is stripped immediately due to ram pressure. The discovery of a possible neutral
hydrogen component of the Sgr stream towards the Galactic anticenter by Putman et al. (2004) suggests that this may
indeed be the last source of star formation fuel for the dwarf that was probably stripped at the pericenter passage
that corresponds in our model to that which occurred about 0.8 Gyr ago. Interestingly, this is also about when the
last significant star formation activity in Sgr took place (Siegel et al. 2007).

The only significant feature of the Sgr core that our model does not reproduce very well is the shape of its
velocity dispersion profile. Although the observed kinematics may still be contaminated, e.g., by Milky Way stars
or tidally stripped material, such
contamination would most probably affect the outer parts, while we also witness some discrepancy in the very center.
While the observed velocity dispersion profile decreases towards the center down to 10 km s$^{-1}$, the simulated one
increases up to about 20 km s$^{-1}$. The observed value is actually very robust and has been established in an
extensive study of the Sgr center by Bellazzini et al. (2008). The velocity dispersion profile of the simulated dwarf
is characteristic of the cuspy, rather concentrated dark matter halo that we use here in the initial conditions. It has
been recently demonstrated by Governato et al. (2010), however,
that the presence of gas and star formation processes can
modify the dark matter halos of isolated dwarfs and cause
them to produce cores. Such cores would manifest themselves in
flatter dispersion profiles of the stars, as observed in Sgr. Mayer et al. (2007) have also shown that the presence of
gas can induce stronger bars, which could further reduce the remnant intrinsic rotation seen in the simulated dwarf.
It seems therefore that including gas and star formation processes may significantly improve the agreement between the
observed and simulated properties of Sgr. However, because only very few hydrodynamical simulations of dwarf galaxies
evolving in the tidal field of their hosts have been performed until now, very little is known about the effect
of additional parameters on the tidal evolution of satellites.
We postpone this much more complex form of study for future work.

The picture of the evolution of the Sgr dwarf we propose in this work fits well within the scenario of the formation
and evolution of dSph galaxies in the Local Group as predicted by the tidal stirring model (Mayer et al. 2001).
This model accounts very well for the observed morphology-density relation for the dwarfs (e.g., Grebel 1999):
those closer to the Milky Way are more affected by tidal forces and have become more spherical, those further away
are still dwarf irregulars. Sgr seems to be in an intermediate stage between objects that have evolved only very little,
like NGC 6822, and those objects that have apparently been substantially evolved by tidal stirring, like the
classical dSph galaxies Draco or Sculptor.
The rather tight orbit of Sgr has allowed it to significantly transform
over a short timescale of 1.3 Gyr and only two pericenter passages. Soon after the recent, second perigalactic passage
Sgr will become more spherical, and more closely resemble a typical, classical dSph galaxy.

On the other hand, the Large Magellanic Cloud (LMC), whose mass is similar to the initial mass we
estimate here for Sgr, has probably just passed its first pericenter around the Milky Way (Besla et al. 2007). Given
that its pericenter distance is rather large, of the order 50 kpc, and its orbit is eccentric,
the LMC core has been little affected by tidal forces.
As shown by Kazantzidis et al. (2010), on such an orbit a satellite disk gets distorted and a
bar can form, but the orbit is too extended and the tidal force therefore too weak to transform the disk into a
spheroid, even over a Hubble time.  That the LMC does in fact presently contain a bar is consistent with this
general picture of dwarf spheroidal satellites being simply the end stage of tidally stirred dwarf disks.  Other
evidence supports the notion of the LMC as a viable prototype for the pre-interaction Sgr progenitor.  Both systems
have very extended star formation histories including recent star formation (e.g., Siegel et al. 2007; Harris \&
Zaritsky 2009). Moreover, Chou et al. (2010) have shown that the LMC and Sgr share very similar chemical enrichment
histories for several $\alpha$ and s-process elements explored, only with Sgr slightly more advanced in its overall
chemical evolution.  The primary difference in the present appearance of these two systems may well be driven primarily
by differences in their current orbital radii and the time they have been bound to the Milky Way.

In this work we have not endeavored to account for the structure and
dynamics of the Sgr tidal arms.  The models that have focused on the
tidal arms of Sgr
(e.g., Ibata et al. 2001; Helmi 2004; Martinez-Delgado et al. 2004; Law et al. 2005; Law \& Majewski 2010) have
tended to attribute more than two perigalacticon passes to Sgr to
account for the length of the arms.  The scenario we propose here,
with just two pericenter passages, does not necessarily mean that we
are in conflict with these other models for the tails. It simply means
that the number of passes that affected the core structure of Sgr
through tidal stirring is most probably limited to two. This however
does not preclude the possibility of previous passes on an orbit of
larger size where the core is unaffected but tails can begin to be
generated. It is obvious that Sgr's orbit must have evolved, both due to dynamical friction and
the growth in mass of the Milky Way (Pe\~narrubia et al. 2006).
More comprehensive models that simultaneously account for both the evolution of the Sgr core and the Sgr tidal debris
in a more realistic, growing Milky Way potential and with an evolving Sgr orbit from infall to the present state
is obviously a goal for future efforts, and one that must be guided by more complete and precise mapping of the
phase space distribution of the extended debris tails.


\acknowledgments

This research was partially
supported by the Polish Ministry of Science and Higher Education under grant NN203025333.
EL{\L} and SRM acknowledge the ASTROSIM network of the European Science Foundation
(Science Meeting 2387) for the financial support of the workshop {\it The Local Universe: from Dwarf Galaxies to Galaxy
Clusters\/} held in Jab{\l}onna near Warsaw in June/July 2009, where this project was initiated.
SK is funded by the Center for Cosmology and Astro-Particle Physics (CCAPP) at The Ohio State University.
EL{\L} is grateful for the hospitality of both the CCAPP and the University of Virginia during her visits.
This work also benefited from an
allocation of computing time from the Ohio Supercomputer Center (http://www.osc.edu/).
SRM acknowledges support from National Science Foundation grant AST-0807945 as well
as the {\it SIM Lite\/} key project {\it Taking Measure of the Milky Way\/} under NASA/JPL contract 1228235.
DRL was supported by NASA through Hubble Fellowship grant \#HF-51244.01 awarded by the Space Telescope
Science Institute, which is operated by the Association of Universities for Research in Astronomy, Inc., for NASA,
under contract NAS 5-26555.
PMF was supported by
an NSF Astronomy and Astrophysics Postdoctoral Fellowship under award AST-0602221,
the NASA Graduate Student Researchers Program, a University of Virginia
Faculty Senate Dissertation-Year Fellowship,
and by the Virginia Space Grant Consortium.


\begin{thebibliography}{}

\bibitem[{Bellazzini et al.}(2008)]{b08} Bellazzini, M., Ibata, R. A., Chapman, S. C., Mackey, A. D., Monaco, L.,
        Irwin, M. J., Martin, N. F., Lewis, G. F., \& Dalessandro, E. 2008, AJ, 136, 1147
\bibitem[{Belokurov et al.}(2006)]{b06} Belokurov, V., et al. 2006, ApJ, 642, L137
\bibitem[{Besla et al.}(2007)]{b07} Besla, G., Kallivayalil, N., Hernquist, L., Robertson, B., Cox, T. J.,
        van der Marel, R. P., \& Alcock, C. 2007, ApJ, 668, 949
\bibitem[{Bullock et al.}(2001)]{b01} Bullock, J. S., Kolatt, T. S., Sigad, Y., Somerville, R. S.,
        Kravtsov, A. V., Klypin, A. A., Primack, J. R., \& Dekel, A. 2001, MNRAS, 321, 559
\bibitem[{Chou et al.}(2010)]{ch10} Chou, M.-Y., Cunha, K., Majewski, S. R., Smith, V. V., Patterson,
        R. J., Martinez-Delgado, D., \& Geisler, D. 2010, ApJ, 708, 1290
\bibitem[{Colpi et al.}(1999)]{co99} Colpi, M., Mayer, L., \& Governato, F. 1999, ApJ, 525, 720
\bibitem[{Correnti et al.}(2010)]{co10} Correnti, M., Bellazzini, M., Ibata, R. A., Ferraro, F. R., \& Varghese, A.
	2010, ApJ, 721, 329
\bibitem[{Fellhauer et al.}(2006)]{f06} Fellhauer, M., et al. 2006, ApJ, 651, 167
\bibitem[{Frinchaboy et al.}(2010)]{f10} Frinchaboy, P. M., Majewski, S. R., Mu\~noz, R. R., \& Patterson, R. J.
        2010, in preparation
\bibitem[{Governato et al.}(2010)]{g10} Governato, F., et al. 2010, Nature, 463, 203
\bibitem[{Grebel et al.}(1999)]{g99} Grebel, E. K. 1999, in IAU Symp. 192, The Stellar Content of Local Group
        Galaxies, ed. P. Whitelock \& R. Cannon (San Francisco: ASP), 17
\bibitem[{Guo et al.}(2009)]{g09} Guo, Y., McIntosh, D. H., Mo, H. J.; Katz, N., van den Bosch, F. C.,
        Weinberg, M., Weinmann, S. M., Pasquali, A., \& Yang, X.  2009, MNRAS, 398, 1129
\bibitem[{Harris \& Zaritsky}(2009)]{hz09} Harris, J., \& Zaritsky, D. 2009, AJ, 138, 1243
\bibitem[{Helmi}(2004)]{helmi04} Helmi, A. 2004, ApJ, 610, L97
\bibitem[{Helmi \& White}(2001)]{hw01} Helmi, A., \& White, S. D. M. 2001, MNRAS, 323, 529
\bibitem[{Hernquist}(1993)]{hern93} Hernquist, L. 1993, ApJS, 86, 389
\bibitem[{Ibata et al.}(1994)]{ibata94} Ibata, R. A., Gilmore, G., \& Irwin, M. J. 1994, Nature, 370, 194
\bibitem[{Ibata et al.}(1997)]{ibata97} Ibata, R. A., Wyse, R. F. G., Gilmore, G., Irwin, M. J., \& Suntzeff, N. B.
        1997, AJ, 113, 634
\bibitem[{Ibata et al.}(2001)]{Ibata2001} Ibata, R., Lewis, G.~F., Irwin, M., Totten, E., \& Quinn, T.\ 2001, ApJ, 551, 294
\bibitem[{Jiang \& Binney}(2000)]{jb00} Jiang, I.-G., \& Binney, J. 2000, MNRAS, 314, 468
\bibitem[{Jimenez et al.}(2003)]{jim03} Jimenez, R., Verde, L., Oh, S. P. 2003, MNRAS, 339, 243
\bibitem[{Johnston et al.}(1995)]{j95} Johnston, K. V., Spergel, D. N., \& Hernquist, L. 1995, ApJ, 451, 598
\bibitem[{Johnston et al.}(1999)]{j99} Johnston, K. V., Majewski, S. R., Siegel, M. H., Reid, I. N., Kunkel, W. E.
        1999, AJ, 118, 1719
\bibitem[{Johnston et al.}(2005)]{j05} Johnston, K. V., Law, D. R., \& Majewski, S. R. 2005, ApJ, 619, 800
\bibitem[{Kazantzidis et al.}(2010)]{k10} Kazantzidis, S., {\L}okas, E. L., Callegari, S., Mayer, L., \& Moustakas, L.
        A., 2010, submitted to ApJ, arXiv:1009.2499
\bibitem[{King}(1966)]{k66} King, I. 1966, AJ, 71, 64
\bibitem[{Klimentowski et al.}(2007)]{k07} Klimentowski, J., {\L}okas, E. L., Kazantzidis, S., Prada, F., Mayer, L.,
        \& Mamon, G. A. 2007, MNRAS, 378, 353
\bibitem[{Klimentowski et al.}(2009)]{k09} Klimentowski, J., {\L}okas, E. L., Kazantzidis, S., Mayer, L., \& Mamon,
        G. A. 2009, MNRAS, 397, 2015
\bibitem[{Klimentowski et al.}(2010)]{k10} Klimentowski, J., {\L}okas, E. L., Knebe, A., Gottl\"ober, S.,
        Martinez-Vaquero, L. A., Yepes, G., \& Hoffman, Y. 2010, MNRAS, 402, 1899
\bibitem[{Klypin et al.}(2002)]{klypin02} Klypin,  A., Zhao,  H., \& Somerville, R. S., 2002, ApJ, 573, 597
\bibitem[{Kunder \& Chaboyer}(2009)]{kc09} Kunder, A., \& Chaboyer, B. 2009, AJ, 137, 4478
\bibitem[{Law et al.}(2005)]{law05} Law, D. R., Johnston, K. V., \& Majewski, S. R. 2005, ApJ, 619, 807
\bibitem[{Law et al.}(2009)]{law09} Law, D. R., Majewski, S. R., \& Johnston, K. V. 2009, ApJ, 703, L67
\bibitem[{Law \& Majewski}(2010)]{lm10} Law, D. R., \& Majewski, S. R. 2010, ApJ, 714, 229
\bibitem[{Lokas}(2002)]{lo02} {\L}okas, E. L. 2002, MNRAS, 333, 697
\bibitem[{Lokas}(2009)]{lo09} {\L}okas, E. L. 2009, MNRAS, 394, L102
\bibitem[{Lokas et al.}(2005)]{lmp} {\L}okas, E. L., Mamon, G. A., \& Prada, F. 2005, MNRAS, 363, 918
\bibitem[{Lokas et al.}(2008)]{lkkm} {\L}okas, E. L., Klimentowski, J., Kazantzidis, S., \& Mayer, L. 2008,
	MNRAS, 390, 625
\bibitem[{Lokas et al.}(2010)]{lkkmc} {\L}okas, E. L., Kazantzidis, S., Klimentowski, J., Mayer, L., \& Callegari, S.
        2010, ApJ, 708, 1032
\bibitem[{Majewski et al.}(2003)]{maj03} Majewski, S. R., Skrutskie, M. F., Weinberg, M. D., \& Ostheimer,
        J. C. 2003, ApJ, 599, 1082
\bibitem[{Majewski et al.}(2004)]{maj04} Majewski, S. R., et al. 2004, AJ, 128, 245
\bibitem[{Martinez-Delgado et al.}(2004)]{martinez04} Martinez-Delgado, D., Gomez-Flechoso, M. A., Aparicio,
        A., Carrera, R. 2004, ApJ, 601, 242
\bibitem[{Mateo et al.}(1998)]{mat98} Mateo, M., Olszewski, E. W., \& Morrison, H. L. 1998, ApJ, 508, L55
\bibitem[{Mayer \& Moore}(2004)]{mm04} Mayer, L., \& Moore, B. 2004, MNRAS, 354, 477
\bibitem[{Mayer et al.}(2001)]{mayer01}  Mayer, L., Governato, F., Colpi, M., Moore, B., Quinn, T., Wadsley, J.,
        Stadel, J., \& Lake, G. 2001, ApJ, 559, 754
\bibitem[{Mayer et al.}(2006)]{mayer06} Mayer, L., Mastropietro, C., Wadsley, J., Stadel, J. \& Moore, B.
        2006, MNRAS, 369, 1021
\bibitem[{Mayer et al.}(2007)]{mayer07} Mayer, L., Kazantzidis, S., Mastropietro, C., \& Wadsley, J. 2007,
        Nature, 445, 738
\bibitem[{McGaugh \& Wolf}(2010)]{mw10} McGaugh, S. S., \& Wolf, J.  2010, ApJ, 722, 248
\bibitem[{Mo et al.}(1998)]{mo98} Mo,  H. J., Mao,  S., \& White,  S. D. M. 1998, MNRAS, 295, 319
\bibitem[{Navarro et al.}(1997)]{nfw97} Navarro, J. F., Frenk, C. S., \& White, S. D. M. 1997, ApJ, 490, 493 (NFW)
\bibitem[{Oh et al.}(2008)]{oh08} Oh, S.-H., de Blok, W. J. G., Walter, F., Brinks, E., \& Kennicutt, R. C.
        2008, AJ, 136, 2761
\bibitem[{Pe\~narrubia et al.}(2006)]{pe06} Pe\~narrubia, J., Benson, A. J., Martínez-Delgado, D., \& Rix, H. W.
        2006, ApJ, 645, 240
\bibitem[{Pe\~narrubia et al.}(2010)]{pe10} Pe\~narrubia, J., Belokurov, V., Evans, W. N., Martinez-Delgado,
        D., Gilmore, G., Irwin, M., Niederste-Ostholt, M., \& Zucker, D. B. 2010, MNRAS, 408, L26
\bibitem[{Putman et al.}(2004)]{p04} Putman, M. E., Thom, C., Gibson, B. K., \& Staveley-Smith, L. 2004, ApJ, 603, L77
\bibitem[{Richstone \& Tremaine}(1986)]{rt86} Richstone, D. O., \& Tremaine, S. 1986, AJ, 92, 72
\bibitem[{Sersic}(1968)]{ser68} S\'ersic, J. L. 1968, Atlas de Galaxies Australes,
        Observatorio Astronomico, Cordoba
\bibitem[{Sawala et al.}(2010)]{s10} Sawala, T., Scannapieco, C., Maio, U., \& White, S. 2010, MNRAS, 402, 1599
\bibitem[{Schombert}(2006)]{schombert06} Schombert, J. M. 2006, AJ, 131, 296
\bibitem[{Siegel et al.}(2007)]{s07} Siegel, M. H., et al. 2007, ApJ, 667, L57
\bibitem[{Stadel}(2001)]{s01} Stadel, J. G. 2001, PhD thesis, Univ. of Washington
\bibitem[{Taffoni et al.}(2003)]{t03} Taffoni, G., Mayer, L., Colpi, M., \& Governato, F. 2003, MNRAS, 341, 434
\bibitem[{Valenzuela et al.}(2007)]{v07} Valenzuela, O., Rhee, G., Klypin, A., Governato, F., Stinson, G.,
        Quinn, T., \& Wadsley, J. 2007, ApJ, 657, 773
\bibitem[{Widrow \& Dubinski}(2005)]{wd05} Widrow, L. M., \& Dubinski, J. 2005, ApJ, 631, 838
\bibitem[{Yanny et al.}(2009)]{y09} Yanny, B., et al. 2009, ApJ, 700, 1282

\end{thebibliography}
\end{document}